\documentclass[aps, showpacs, showkeys,nofootinbib,floatfix]{revtex4}

\usepackage{amssymb}
\usepackage{amsmath}
\usepackage{graphicx}

\begin{document}

\title{Constraints on kinematic models from the latest observational data}

\author{Jianbo Lu}
\email{lvjianbo819@163.com}
 \affiliation{Department of Physics, Liaoning Normal University, Dalian 116029, P. R. China}
\author{Lixin Xu}
 \affiliation{Korea Astronomy and Space Science Institute, Daejon  305-348,
  Korea}
 \affiliation{School of Physics and Optoelectronic Technology, Dalian University of Technology, Dalian, 116024, P. R. China}
 \author{Molin Liu}
 \affiliation{College of Physics and Electronic Engineering, Xinyang Normal University, Xinyang, 464000, P. R. China}

\begin{abstract}
Kinematical models are constrained by  the latest observational data
from geometry-distance measurements, which include 557 type Ia
supernovae (SNIa) Union2 data
 and 15 observational Hubble data.
 Considering two parameterized
deceleration parameter, the values of current deceleration parameter
$q_{0}$, jerk parameter  $j_{0}$ and transition redshift $z_{T}$,
are obtained. Furthermore, we show the departures for two
parameterized kinematical models from $\Lambda$CDM model according
to the evolutions of  jerk parameter $j(z)$. Also, it is shown that
the constraint on jerk parameter $j(z)$ is weak by the current
geometrical observed data.

\end{abstract}


\maketitle

\section{$\text{Introduction}$}
The recently cosmic observations
\cite{SNeRiess}\cite{CMBSpergel}\cite{LSSPope} suggest that the
expansion of present universe is speeding up. The accelerated
expansion of the universe is usually  attributed to the fact that
dark energy (DE) is an exotic component with negative pressure, such
as cosmological constant, quintessence \cite{quintessence}, phantom
\cite{phantom}, quintom \cite{quintom}, generalized Chaplygin gas
\cite{GCG}, agegraphic dark energy \cite{agegraphic}, etc
\cite{other-models}. Also, the accelerating universe is related to
the modification of the gravity theory at large scale, such as
$f(R)$ modified gravity theory \cite{fR} and higher dimensional
theory \cite{higherD}, etc. These approaches correspond to the
dynamics of the universe. For more information about dynamics of
universe, please see review papers \cite{review}.

Another route is the kinematical approach, which holds true
regardless of the underlying cosmic dynamics
\cite{kinematicalmodel}, i.e., it depends neither on the validity of
any particular metric theory of gravity nor on the matter-energy
content of the observed universe \cite{09043550kinematical}. It is
only related to the weaker assumption that space-time is homogeneous
and isotropic so that the FRW metric is still valid
\cite{09043550kinematical}. Then the kinematic approach is also
called cosmokinetics \cite{cosmokineticas}, cosmography
\cite{cosmogrphy}, or Friedmannless cosmology \cite{friedmannless}.
For kinematic equations, there are Hubble parameter
$H=\frac{\dot{a}}{a}$, deceleration parameter
$q=-a\ddot{a}/\dot{a}^{2}$, and jerk parameter
$j=-\dddot{a}a^{3}/(a\dot{a}^{3})$. It can be seen that parameters
$H$, $q$ and $j$ are purely kinematical, since they are independent
of any gravity theory, and all of them are only related to scale
factor $a$ (or redshift $z$, since $a=\frac{1}{1+z}$).

The benefit of the kinematical analysis is that it
 has the fewer assumptions and  a different set of models are explored
  for comparison with dynamical scenario. Since the origin of cosmic acceleration is unknown,  the choice
  of
parameterized kinematical model  is essentially arbitrary. But
inappropriate kinematical model  could imply an unphysical universe
at earlier time. For instance, for the model $q(z)=q_{0}+q_{1}z$, it
has $q(z)>1/2$ at high redshift \cite{q0.5},
 which is not consistent with the matter dominated universe\footnote{For the
  matter dominated universe, it has $q(z)=1/2$, then there  must be $q(z) \leq
1/2$ for any kinematical or dynamical model.}.    The reason is
simple that the model  $q(z)=q_{0}+q_{1}z$, a expansion of $q(z)$ at
low redshift, is not reliable at high redshift. As a complementarity
to dynamical approach, in this paper we constrain two parameterized
kinematical models by using the latest observational data: 557 type
Ia supernovae (SNIa) Union2 dataset
 and 15 observational Hubble data.

\section{$\text{The kinematical approach and models}$}

The dimensionless Hubble and deceleration parameters are defined by
the first and  second derivative of scale factor
\begin{equation}
H\equiv \frac{\dot{a}}{a}=-\frac{1}{1+z}\frac{dz}{dt},\label{e1}
\end{equation}
\begin{equation}
q \equiv
-\frac{1}{H^{2}}(\frac{\ddot{a}}{a})=\frac{1}{2}(1+z)\frac{[H(z)^{2}]^{'}}{H(z)^{2}}-1,\label{e2}
\end{equation}
 where  $^{"}$dot$^{"}$ denotes the
derivative with respect to cosmic time $t$. $H$ and $q$ describe the
rate of expansion and acceleration of universe. The relation between
$H$ and $q$ is written as
\begin{equation}
H=H_{0} \exp[\int^{z}_{0}[1+q(u)] d\ln(1+u)],\label{e3}
\end{equation}
$H_{0}$ is  Hubble constant. In this paper, subscript "0" denotes
the current value of cosmological quantities.

 Similar, the jerk
parameter $j$ is defined as the dimensionless third derivative of
 scale factor with respect to cosmic time
\begin{equation}
j \equiv
-\frac{1}{H^{3}}(\frac{\dot{\ddot{a}}}{a})=-[\frac{1}{2}(1+z)^{2}\frac{[H(z)^{2}]^{''}}{H(z)^{2}}-(1+z)\frac{[H(z)^{2}]^{'}}{H(z)^{2}}+1].\label{e4}
\end{equation}
 The use of the cosmic jerk parameter provides a more natural
parameter space for kinematical studies, and transitions between
phases of different cosmic acceleration are more naturally described
by models incorporating a cosmic jerk \cite{jerkvalue}. Especially,
since for flat $\Lambda$CDM model it has a constant jerk with $j(z)
= -1$, jerk parameter can provides us with a simple, convenient
approach to search for departures  from the cosmic concordance
model, $\Lambda$CDM, just as deviations from $w = -1$ done in more
standard dynamical analyses. The  jerk parameter is related with the
deceleration parameter by the following differential equation
\begin{equation}
j=-[q+2q^{2}+(1+z)\frac{dq}{dz}].\label{e5}
\end{equation}

From Eqs. (\ref{e3}) and (\ref{e5}), we can see that  the
expressions of Hubble parameter and jerk parameter can be given by
  deceleration parameter. Here we consider two
parameterized deceleration parameters,
$q(z)=q_{0}+\frac{q_{1}z}{1+z}$ ($M_{1}$) \cite{qa} and
$q(z)=\frac{1}{2}+\frac{q_{1}+q_{2}z}{(1+z)^{2}}$ ($M_{2}$)
\cite{qgong}. $M_{1}$  is one order expansion of scale factor $a$ at
present $(a = 1)$, i.e., $q(a) = q_{0} + q_{1}(1-a)$. $M_{2}$ is an
alternative parametrization to the three-epoch model\footnote{The
three-epoch model \cite{threeepoch} is a alternative scenario for
the model $q(z)=q_{0}+q_{1}z$, but the problem is that this function
$q(z)$ is not smooth \cite{qgong}.}. For these two models, the
expressions of Hubble parameter, deceleration parameter and jerk
parameter are shown in Table \ref{table1}.

For other parameterized deceleration parameter appeared  in Ref.
\cite{09043550kinematical}, such as model $q(z)=q_{0}+q_{1}z$ and
$q(z)=q_{0}$=constant, we will not discuss: since the former model
is only interested at low redshift, not all the observed data from
557 SNIa Union2  dataset (redshift interval $0.015\leq z \leq 1.4$)
and 15 observational Hubble data (redshift interval $0 \leq z \leq
1.75$)   can be used  to constrain
  its  evolution; though the latter one indicates
an accelerating universe \cite{09043550kinematical}, it does not
describe a transition of universe from decelerated expansion to
accelerated expansion. Furthermore, for two-epoch model ($q=q_{0}$,
$z\leq z_{T}$; $q=q_{1}$, $z > z_{T}$), the $q(z)$ function is not
smooth, so it will not appeared in the paper, too.
\begin{table}
\vspace*{-12pt}
\begin{center}
\begin{tabular}{c | c| c  }
\hline\hline  $H(z)$ &  $q(z)$&  $j(z)$
\\\hline
  $H_{0}(1+z)^{1+q_{0}+q_{1}}\exp(-\frac{q_{1}z}{1+z})$   & $q_{0}+\frac{q_{1}z}{1+z}$
   & $-q_{0}-\frac{q_{1}z}{1+z}-2(q_{0}+\frac{q_{1}z}{1+z})-(1+z)[\frac{q_{1}}{1+z}-\frac{q_{1}z}{(1+z)^{2}}]$
 \\\hline
$H_{0}(1+z)^{\frac{3}{2}}
\exp[\frac{2q_{1}z+(q_{1}+q_{2})z^{2}}{2(1+z)^{2}}]$
  & $\frac{1}{2}+\frac{q_{1}+q_{2}z}{(1+z)^{2}}$
&$-\frac{1}{2}-\frac{q_{1}+q_{2}z}{(1+z)^{2}}-2(\frac{1}{2}+\frac{q_{1}+q_{2}z}{(1+z)^{2}})^{2}-(1+z)[\frac{q_{2}}{(1+z)^{2}}-\frac{2(q_{1}+q_{2}z)}{(1+z)^{3}}]$
  \\\hline\hline
       \end{tabular}
       \end{center}
        \caption{The expressions of Hubble parameter $H(z)$, deceleration parameter $q(z)$ and jerk parameter $j(z)$ for two models.}\label{table1}
       \end{table}

\section{$\text{Data and analysis Methods}$}

Since SNIa behave as excellent standard candles, they can be used to
directly measure the expansion rate of the universe up to high
redshift for comparison with the present rate.
 Theoretical  cosmic parameters are
determined by minimizing the quantity \cite{chi2SNe}
\begin{equation}
\chi^{2}_{SNIa}(H_{0}, \theta)=\sum_{i=1}^{N}\frac{(\mu_{obs}(z_{i})
-\mu_{th}(z_{i},H_{0},\theta))^2}{\sigma^2_{obs;i}},\label{e6}
\end{equation}
where $N=557$ for SNIa Union2  data; $\sigma^2_{obs;i}$ are errors
due to flux uncertainties, intrinsic dispersion of SNIa absolute
magnitude and peculiar velocity dispersion, respectively; $\theta$
denotes the model parameters; $\mu_{obs}$ is the observed values of
distance modulus and can be given by the SNIa dataset; the
expression of theoretical distance modulus $\mu_{th}$ is related to
the apparent magnitude of
  supernova at peak brightness $m$ and the absolute magnitude $M$,
\begin{equation}
\mu_{th}(z_{i})\equiv
m_{th}(z_{i})-M=5log_{10}(D_{L}(z))+\mu_{0},\label{e7}
\end{equation}
where luminosity distance
\begin{equation}
D_{L}(z)=H_{0}d_{L}(z)=(1+z)\int_{0}^{z}\frac{H_{0}dz^{'}}{H(z^{'};\theta)},\label{e8}
\end{equation}
and
\begin{equation}
\mu_{0}=5log_{10}(\frac{H_{0}^{-1}}{Mpc})+25=42.38-5log_{10}h.\label{e9}
\end{equation}
It should be noted that $\mu_{0}$ is independent of the data and the
dataset, though it is a nuisance parameter. By expanding the
$\chi^{2}$ of Eq. (\ref{e6}) relative to $\mu_{0}$,  the
minimization with respect to $\mu_{0}$ can be made trivially
\cite{SNeABC}\cite{chi2sneli}
\begin{equation}
\chi^{2}_{SNIa}(\theta)=A(\theta)-2\mu_{0}B(\theta)+\mu_{0}^{2}C,\label{e10}
\end{equation}
where
\begin{equation}
A(\theta)=\sum_{i=1}^{N}\frac{[\mu_{obs}(z_{i})-\mu_{th}(z_{i};\mu_{0}=0,\theta)]^{2}}{\sigma_{i}^{2}},\label{e11}
\end{equation}
\begin{equation}
B(\theta)=\sum_{i=1}^{N}\frac{\mu_{obs}(z_{i})-\mu_{th}(z_{i};\mu_{0}=0,\theta)}{\sigma_{i}^{2}},\label{e12}
\end{equation}
\begin{equation}
C=\sum_{i=1}^{N}\frac{1}{\sigma_{i}^{2}},\label{e13}
\end{equation}
Evidently, Eq. (\ref{e6}) has a minimum for $\mu_{0}=B/C$ at
\begin{equation}
\widetilde{\chi}^{2}_{SNIa}(\theta)=A(\theta)-B(\theta)^{2}/C.\label{e14}
\end{equation}
 Since $\chi^{2}_{SNIa,min}=\widetilde{\chi}^{2}_{SNIa,min}$ and $\widetilde{\chi}^{2}_{SNIa}$ is independent of nuisance
parameter $\mu_{0}$, here we utilize
  expression (\ref{e14}) to displace (\ref{e6}) for SNIa
constraint. Alternatively, one can also perform  a uniform
marginalization over the nuisance parameter $\mu_{0}$ thus obtaining
\cite{chi2sne-ln-refs1,chi2sne-ln-refs2,chi2sne-ln-refs3}
\begin{equation}
\chi^{2}_{SNIa}(\theta)=A(\theta)-\frac{B(\theta)^{2}}{C}+\ln(\frac{C}{2\pi}).\label{chi2SNIa-ln}
\end{equation}
Comparing Eq. (14) with Eq. (15), it can be seen that two
$\chi^{2}$s are equivalent for using them to constrain cosmological
models, since they are only different from a constant term
$\ln(\frac{C}{2\pi})$, i.e. if one marginalize over all values of
$\mu_{0}$, as in Ref. \cite{chi2sne-marginalization}, that would
just add a constant and would not change the constraint results.

Recently, Stern $et$ $al$ obtained the Hubble parameter $H(z)$ at 12
different redshifts from the differential ages of passively evolving
galaxies \cite{12Hubbledata}.  And in  Ref. \cite{3Hubbledata},
authors obtained $H(z = 0.24) = 79.69 \pm 2.32$, $H(z = 0.34) = 83.8
\pm 2.96$, and $H(z = 0.43) = 86.45 \pm 3.27$ by taking the BAO
scale as a standard ruler in the radial direction.
 Using these data we can constrain cosmological
models by minimizing
\begin{equation}
\chi^2_{Hub}(H_{0},
\theta)=\sum_{i=1}^{N}\frac{\left[H_{th}(z_i)-H_{obs}(H_{0},\theta,
z_i)\right]^2}{\sigma^2_{obs;i}},\label{e15}
\end{equation}
where $H_{th}$ is the predicted value for the Hubble parameter,
$H_{obs}$ is the observed value, $\sigma^2_{obs;i}$ is the standard
deviation measurement uncertainty. Here the nuisance parameter
$H_{0}$ is marginalized in the following calculation
 with a Gaussian prior, $H_{0} = 74.2 \pm 3.6$ km~s$^{-1}$ Mpc$^{-1}$ \cite{H0prior}.

\begin{figure}[!htbp]
  \includegraphics[width=5cm]{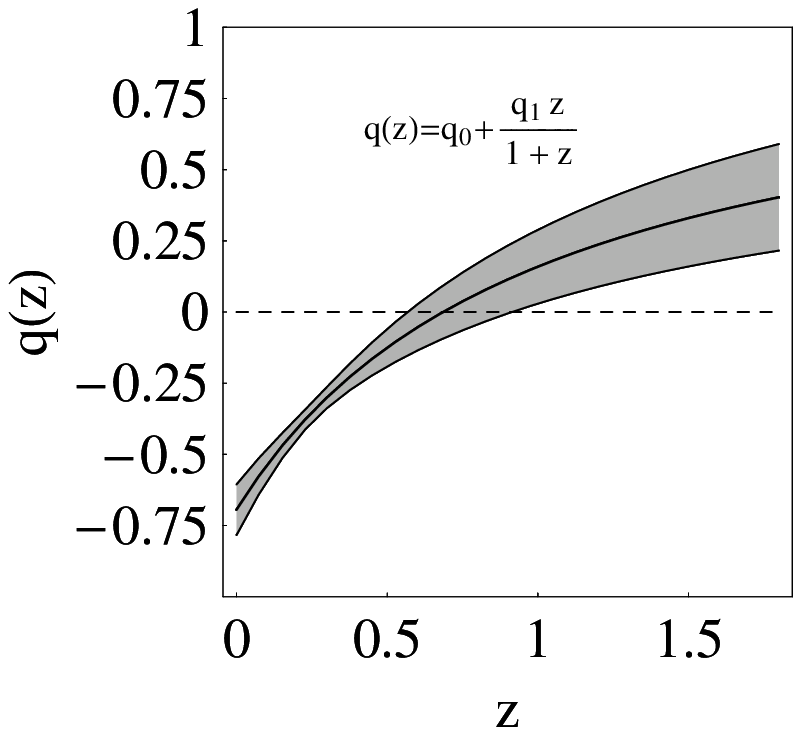}
~~~~~~\includegraphics[width=5cm]{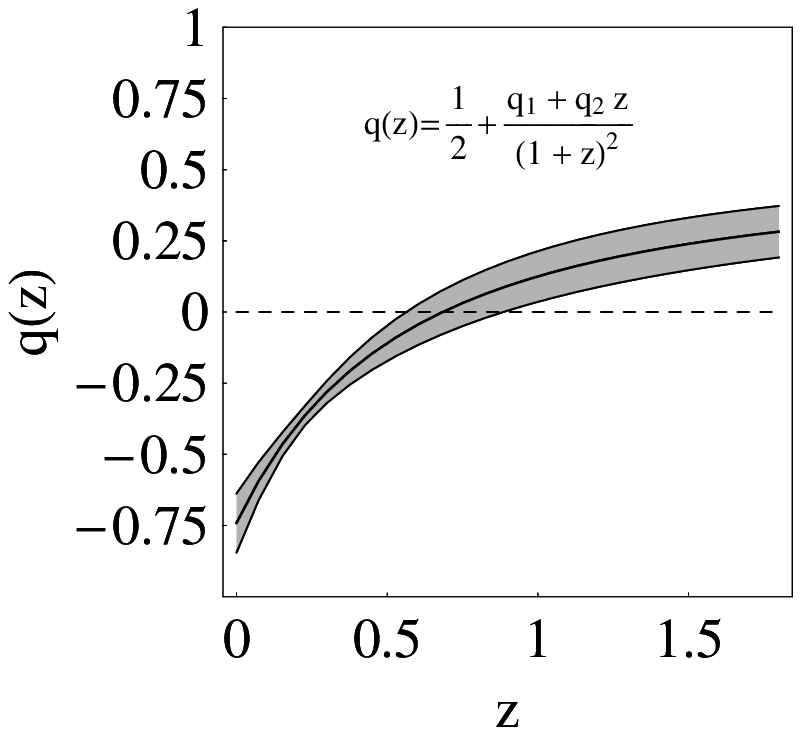}
~~~~~~\includegraphics[width=5cm]{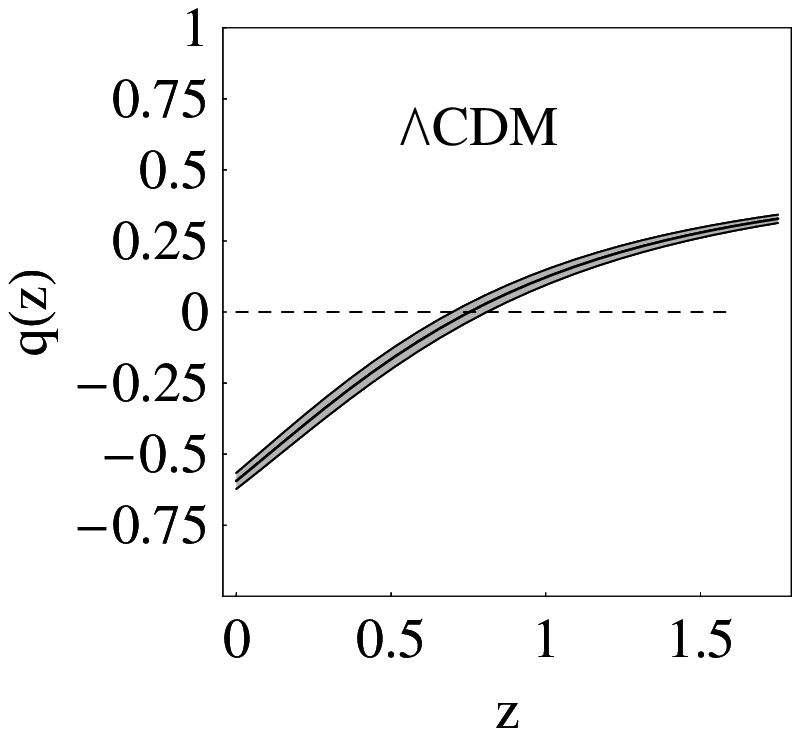}\\
  \caption{The best-fit evolutions of $q(z)$ with $1\sigma$ confidence level constrained from
  557 SNIa Union2 dataset and 15 observational Hubble data.}\label{figureq}
\end{figure}

The likelihood function is written as $L\propto e^{-\chi^{2}/2}$,
and the total $\chi^{2}$ equals,
\begin{equation}
\chi^{2}_{total}=\tilde{\chi}^{2}_{SNIa}+\chi^{2}_{Hub},\label{e16}
\end{equation}
where $\tilde{\chi}^{2}_{SNIa}$ and $\chi^{2}_{Hub}$ are the ones
described in Eq. (\ref{e14}) and Eq.  (\ref{e15}), respectively. It
is easy to see that the matter density $\Omega_{m}$ are not
contained explicitly in the $\chi^{2}_{total}$. Then, the constraint
results may not depend on the dynamic variables $\Omega_{m}$, and
gravitation theory.

 \begin{table}
 \vspace*{-12pt}
 \begin{center}
 \begin{tabular}{c | c| c | c| c|c  } \hline\hline model & $\chi_{min}^{2}$ & $\chi_{min}^{2}$/dof& $q_{0}$ $(1\sigma)$ & $j_{0}$ $(1\sigma)$ & $z_{T}$ $(1\sigma)$ \\\hline
 $M_{1}$ & 554.335 (478.060) & 0.973(1.163)     &$-0.701^{+0.089}_{-0.089}$ ($-0.653^{+0.092}_{-0.093}$)
  &$-2.000^{+0.443}_{-0.442}$ ($-1.823^{+0.456}_{-0.457}$)  & $0.689^{+0.227}_{-0.117}$ ($0.674^{+0.257}_{-0.124}$) \\\hline
 $M_{2}$   & 554.560 (478.334) &   0.973(1.167)     &$-0.749^{+0.103}_{-0.103}$  ($-0.698^{+0.108}_{-0.108}$)
 & $-2.619^{+0.602}_{-0.602}$ ($-2.386^{+0.619}_{-0.620}$) & $0.687^{+0.198}_{-0.121}$ ($0.667^{+0.215}_{-0.126}$)  \\\hline
 $\Lambda$CDM & 557.008 (479.283) &   0.975(1.166)     &$-0.598^{+0.028}_{-0.027}$ ($-0.575^{+0.030}_{-0.030}$)
  &$-1$  & $0.761^{+0.055}_{-0.055}$ ($0.716^{+0.056}_{-0.056}$) \\\hline\hline
 \end{tabular}
 \end{center}
 \caption{The values of the current deceleration parameter $q_{0}$, jerk parameter $j_{0}$, and
  transition redshift $z_{T}$  against the model, obtained by using 557 SNIa Union2 data and 15  observational Hubble data
  (the numerical results in brackets  correspond to the constraints from 397 SNIa Constitution data and 15  observational Hubble data).}\label{table2}
 \end{table}

\begin{figure}[!htbp]
  \includegraphics[width=5.5cm]{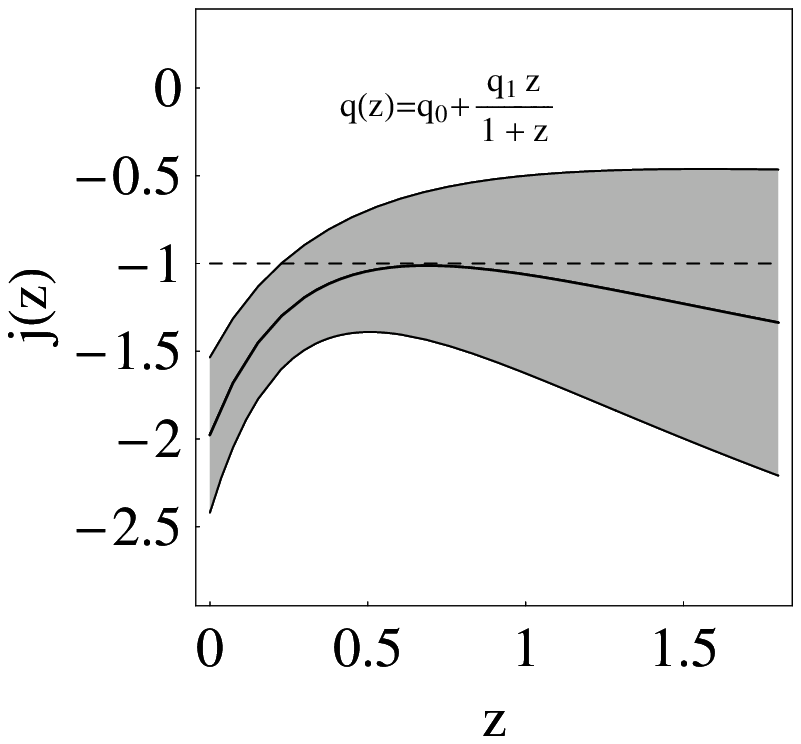}
 ~~~~~~\includegraphics[width=5.5cm]{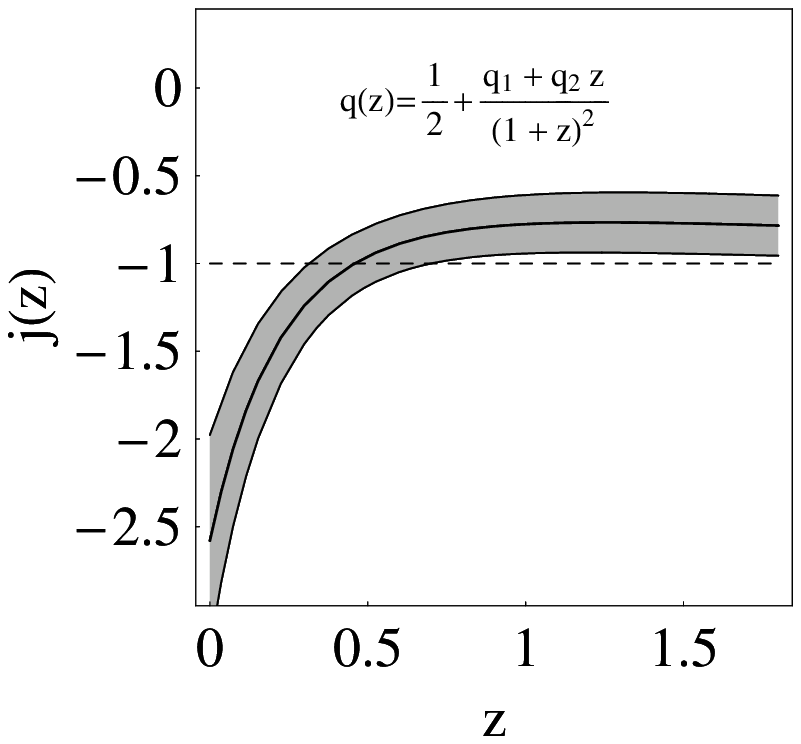}\\
  \caption{The best-fit evolutions of $j(z)$ with $1\sigma$  confidence level  constrained from
  557 SNIa Union2 dataset and 15 observational Hubble data.}\label{figurej}
\end{figure}

The evolutions of two kinematical   parameters, $q(z)$ and  $j(z)$,
are plotted in Fig. \ref{figureq} and \ref{figurej} for two models,
respectively. According to these figures, constraint results of
cosmic parameters are obtained and shown in table \ref{table2}, with
the constraints  on model parameters  shown in table
\ref{table-modelparams}. Relative to the evolutions of deceleration
parameters $q(z)$ it can be seen that the constraints on jerk
parameters $j(z)$, defined by the third derivative with respect to
scale factor $a$, are  weaker for these two models from the
distance-measurement data. Furthermore, in Fig. \ref{figureq} as a
contrast we also plot the evolution of deceleration parameter $q(z)$
for $\Lambda$CDM model by the current geometry-distance observed
data (and for this model, jerk parameter $j(z)=-1$ almost). From
table \ref{table2}, it is shown that for the dynamical $\Lambda$CDM
model, the constraint results tend to favor the bigger values of
current deceleration parameter $q_{0}$ and transition redshift
$z_{T}$ relative to the cases of two kinematical models.

 \begin{table}
 \vspace*{-12pt}
 \begin{center}
 \begin{tabular}{c |  c  } \hline\hline model & The values of model parameters $(1\sigma)$ \\\hline
   $M_{1}$       &   $q_{0}=-0.701^{+0.089}_{-0.089}$ ($q_{0}=-0.653^{+0.092}_{-0.093}$), ~~
      $q_{1}=1.718^{+0.475}_{-0.504} $ ($q_{1}=1.623^{+0.493}_{-0.529} $)   \\\hline
    $M_{2}$    &   $q_{1}=-0.253^{+0.653}_{-0.673} $ ($q_{1}=-0.286^{+0.692}_{-0.714} $), ~~
      $q_{2}=-1.249^{+0.106}_{-0.158} $ ($q_{2}=-1.198^{+0.165}_{-0.162} $) \\\hline
 $\Lambda$CDM           &  $\Omega_{0m}=0.268^{+0.019}_{-0.018}$ ($\Omega_{0m}=0.283^{+0.021}_{-0.019}) $  \\\hline\hline
 \end{tabular}
 \end{center}
 \caption{The $1\sigma$ confidence level of model parameters for the models: $q(z)=q_{0}+\frac{q_{1}z}{1+z}$, $q(z)=\frac{1}{2}+\frac{q_{1}+q_{2}z}{(1+z)^{2}}$
 and $\Lambda$CDM model by using the  557 SNIa Union2 data and 15  observational Hubble data (the numerical results in brackets
  correspond to the constraints from 397 SNIa Constitution data and 15  observational Hubble data).}\label{table-modelparams}
 \end{table}

\section{$\text{Conclusions}$}
In this paper, kinematic models are constrained by the latest
observational data: 557 SNIa Union2 dataset and 15 observational
Hubble data. Generally, the expansion rhythm of current universe
$q_{0}$ and transitional time from decelerated expansion to
accelerated expansion $z_{T}$, depends on the parameterized form of
kinematical equations. Here we consider two parameterized
deceleration parameter. The best fit values of
 $z_{T}$, $q_{0}$ and $j_{0}$ with 1$\sigma$ confidence
level are obtained. From table \ref{table2} we can see that the
values of $z_{T}$ indicated by these two models much approach each
other. From  Fig. \ref{figureq} and \ref{figurej} it can be seen
that for the two kinematical models, the constraints on jerk
parameters $j(z)$ are weak by the current observed data. In
addition, we also can see the deviation of jerk parameter from $j =
-1$ according to the Fig. \ref{figurej}, with measuring the
departures for kinematical models from $\Lambda$CDM model.
Furthermore, considering Refs.
\cite{09043550kinematical}\cite{jerkvalue}\cite{q0.5xu}, where most
models indicate that current data favors $j_{0} < -1$ case, which is
consistent with our results. At last, for comparing the differences
of constraint results  on cosmic parameters between the different
SNIa data, we also consider the case of displacing the 557 SNIa
Union2 data\cite{557Union2}\footnote{The
 SNIa Union2 data are obtained, by adding  new datapoints (including
the high redshift SNIa) to the SNIa Union \cite{307Union} data,
making a number of refinements to the Union analysis chain,
refitting all light curves with the SALT2 fitter.} with 397  SNIa
Constitution data \cite{397Constitution}\footnote{The 397
Constitution data are obtained by adding 90 SNIa from CfA3 sample to
307 SNIa Union sample\cite{307Union}. CfA3 sample are all from the
low-redshift SNIa, $z<0.08$, and these 90 SNIa are calculated with
using the same Union cuts.} in the above combined constraints, and
the latter constraint results are listed  in brackets in table
\ref{table2}. According to  table  \ref{table2}, it seems that the
constraint results favor a bigger value of current deceleration
parameter $q_{0}$ and jerk parameter $j_{0}$, and a smaller value of
transition redshift $z_{T}$.

\textbf{\ Acknowledgments } The research work is supported by the
 National Natural Science Foundation  (Grant No. 10875056), NSF (10703001) and NSF (No.11005088)  of P.R.
  China.

\end{document}